 \documentclass[final,5p,times,twocolumn,authoryear]{elsarticle}

\usepackage{amssymb}
\usepackage{lipsum}

\begin{document}

\begin{frontmatter}



\title{Enhanced plasma ion heating  by lasers in inhomogeneous external magnetic field}

\author{Rohit Juneja$^{1*}$} 
\ead{onlyforjuneja@gmail.com}
\author{Trishul Dhalia$^{1}$}
\author{Amita Das$^{1*}$} 
\ead{amita@iitd.ac.in}

\address{$^{1}$Department of Physics, Indian Institute of Technology Delhi, Hauz Khas, New Delhi 110016, India \\}

\begin{abstract}
Recent studies have shown direct ion heating (\cite{vashistha2020new,Juneja_2023}) by lasers EM (Electromagnetic) wave interacting with a plasma threaded by an external uniform magnetic field. The EM wave frequency was near the lower hybrid (LH) resonance frequency.  The   LH resonance occurs at the edge of the pass band of the magnetized dispersion relation.  The group speed of the wave is negligible at resonance. In these studies, the energy absorption remains essentially confined at the plasma surface. However, to heat the ions in the bulk plasma and at a desired location, a tailored inhomogeneous external magnetic field profile has been chosen here. The strength of the magnetic field at the plasma edge is such that the EM wave frequency lies inside the pass band, where the group velocity has a significant value. It enables the wave to enter the bulk plasma. The external magnetic field is then spatially tailored appropriately to have the LH resonance at a desired spatial location inside the plasma. The Particle-In-Cell (PIC) simulations using the OSIRIS4.0 platform have been carried out, which demonstrates that the EM wave pulse comes to a standstill at the location of the resonance. The wave pulse is observed to break down subsequently, and the energy consequently goes dominantly to the local plasma ions. The absorption is significantly enhanced compared to the case in which the magnetic field profile was homogeneous.  The dependence of absorption on the choice of magnetic field profile, the laser intensity, etc., has also been carried out.

\end{abstract}




\end{frontmatter}



\section{Introduction}
\label{sec:Introduction}
Laser energy absorption by plasma is crucial for a variety of applications. There are various mechanisms by which plasma heating can be obtained by a laser pulse (\cite{kaw1969laser,kaw2017nonlinear,das2020laser,yabuuchi2009evidence}). In some applications such as fusion, medical, etc.,  (\cite{borghesi2010progress, roth2001fast,  atzeni2002first, bulanov2002feasibility, naumova2009hole, steward1973proton}) energetic ions are of paramount importance. Furthermore, there is always an interest in seeking efficient schemes for the creation of energetic ions. (\cite{esirkepov2004highly, kumar2019excitation, turrell2015ultrafast, pfotenhauer2010cascaded, macchi2013ion, henig2009radiation, robson2007scaling, snavely2000intense, fuchs2006laser, robinson2008radiation, haberberger2012collisionless, wilks2001energetic}). In un-magnetized plasma, the energetic ions are created by a two-step process wherein the laser dumps its energy into the lighter electron species, which subsequently gets transferred to ions by collisional and/or collective processes (\cite{Kruer1985JBHB, Brunel1987NotsoresonantRA, ping2008absorption,  wilks1992absorption, chopineau2019identification, freidberg1972resonant, stix1965radiation, gibbon1992collisionless}). Schemes like Radiation pressure acceleration (RPA) (\cite{wilks1992absorption, esirkepov2004highly, macchi2013ion}) and Target normal sheath acceleration (TNSA)(\cite{wilks2001energetic, snavely2000intense}) are also employed for ion acceleration. Recently, our group has proposed novel schemes (\cite{vashistha2020new,Juneja_2023}) of ion heating by applying an external magnetic field. The magnetic field strength was chosen to be appropriate so as to have a magnetized response from the lighter electron species in a laser cycle. The ions behaved as un-magnetized fluid as their gyrofrequency was much smaller than the laser cycle.  The required strength of the magnetic field to achieve this is quite high.  However, technological developments towards increasing the strength of magnetic fields (\cite{nakamura2018record, korneev2015gigagauss}) along with the arrival of long-wavelength intense lasers (e.g., CO$_2$ lasers), very soon, this regime will be experimentally realizable. This has led to a lot of research interest lately in this particular regime (\cite{maity2022mode,mandal2021electromagnetic,dhalia2023harmonic,goswami2022observations}). In this study, we demonstrate using Particle-In-Cell (PIC) simulations that by appropriate spatial tailoring of the applied external magnetic field, the efficiency of energy transfer from the laser to the ions can be significantly enhanced. Furthermore, the technique also helps in heating the ions at a localized region of choice. This is often desirable in the context of several applications. This paper contains the following sections. In section \ref{sim}, simulation details, along with the choice of various simulation parameters, have been provided. In section \ref{sec:ResultDiscussions}, simulation observations have been presented. In particular, the influence of the inhomogeneous magnetic field on EM wave propagation inside the bulk plasma and the enhanced ion heating at the LH resonance layer in the plasma is discussed. The influence of the spatial variation of the magnetic field profile on ion heating has also been studied.  A comparison between the non-relativistic and relativistic intensity of the laser pulse has been provided. Finally, in section \ref{conclusion}, we summarize our findings.

\begin{figure}
  \centering
  \includegraphics[scale = 0.22]{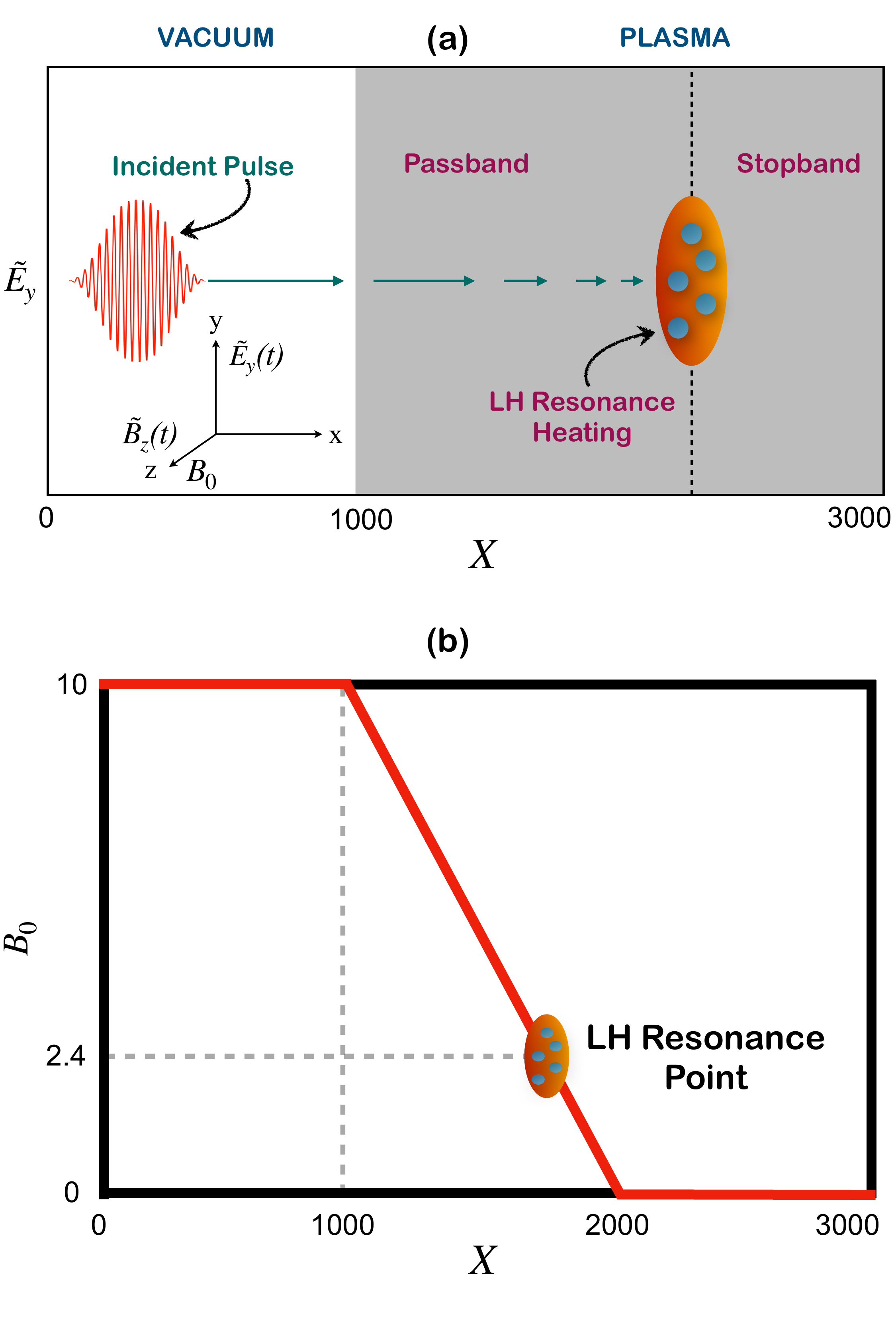}
  \caption{Schematic (not to scale) showing that in (a) Interacting Laser pulse going inside the plasma and get absorbed at the LH resonance layer and in (b) Space variation of the external magnetic field.}
\label{Fig:schematic}
\end{figure}

\begin{table*}
    \caption{Simulation parameters in normalized units and possible values in standard units.} 
  \begin{center}

\setlength{\tabcolsep}{14pt}
  \begin{tabular}{ccc}
  \hline
  \hline
  \hline
      \textbf{Parameters}  & \textbf{Normalized value}   &   {\textbf{Value in standard units}}\\
      \hline
      \hline
      \hline\\
      \hline
    {\textbf{Laser Parameters}} \\
    \hline\\
      Frequency ($\omega_{L}$)  & $0.2 \omega_{pe} $ & $ 0.2 \times 10^{15} Hz$\\
      Wavelength ($\lambda_{L}$)  & 31.4$c/\omega_{pe}$ &9.42 $\mu m$\\
      Intensity ($I_0$)  & $a_0 = 0.5$ & $3.85\times10^{15}Wcm^{-2}$\\\\
      \hline
    {\textbf{Plasma Parameters}} \\
    \hline\\
      Number density ($n_{0}$) & 1 & $3.15 \times 10^{20} cm^{-3}$\\
       Electron plasma frequency ($\omega_{pe}$) & 1 & $10^{15} Hz$\\
       Skin depth ($c/\omega_{pe}$) & $1$ & $0.3{\mu}m$\\\\
       \hline
       {\textbf{Simulation Parameters}} \\
       \hline\\
       $L_x$ & $3000$ & $900 \mu m$\\
       $dx$ & $0.05$ & $15 nm$\\
       $dt$ & $0.02$ & $ 2 \times 10^{-17} s$\\ \\
      \hline
      \hline
      \hline
  \end{tabular}
  \end{center}
  \label{Table}
\end{table*}

\section{Simulation details}
\label{sim}

\begin{figure}
  \centering
  \includegraphics[scale = 0.2]{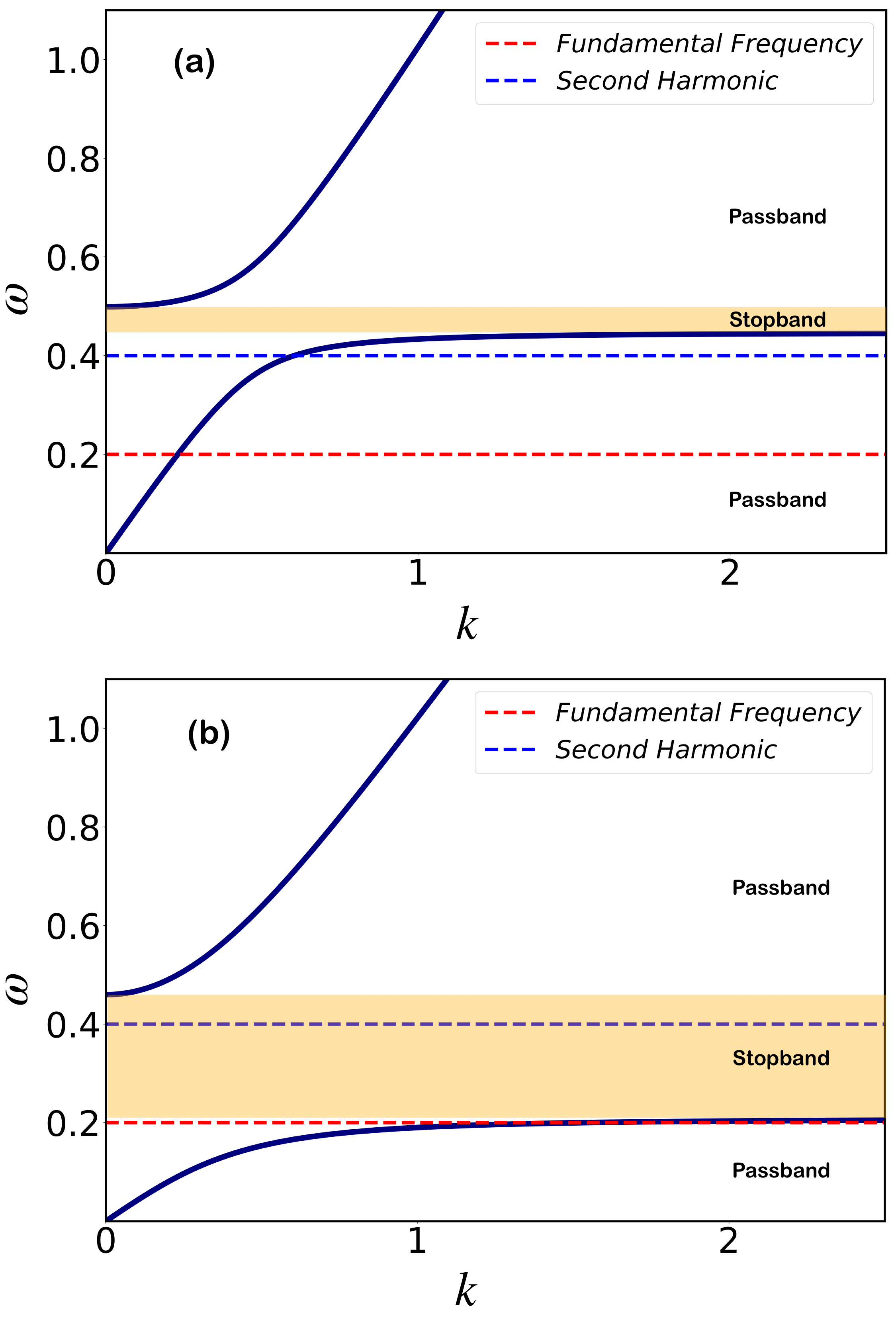}
  \caption{Dispersion curves in X-mode geometry zoomed near LH resonance showing different passbands and stopbands for (a) $B_0~=~10$ and (b) $B_0~=~2.4$.}
\label{Fig:dispersion_curve}
\end{figure}

This study employed a massively parallel Particle-In-Cell code, OSIRIS 4.0 (\cite{hemker2000particle, fonseca2002osiris, fonseca2008one}), to conduct one-dimensional (1D) Particle-In-Cell (PIC) simulations. A one-dimensional simulation box with a length of $L_x = 3000 c/\omega_{pe}$ has been considered. Here, $c$ denotes the speed of light in a vacuum, while $\omega_{pe}$ represents the electron-plasma frequency. In all our simulations the time has been normalized by  $t_{n}$ = ${\omega_{pe}}^{-1}$. For length, the skin depth  $x_{n} = c/\omega_{pe}$ has been chosen. Normalization of the electric and magnetic fields is done by $E_{n} = B_{n} = m_{e}c\omega_{pe}/e$, where $m_{e}$ is the mass of the electron and $e$ represents the magnitude of the electronic charge. Absorbing boundary conditions have been taken for both fields and particles in both directions. We have considered $60000$ grid points (cells) in our simulations, which correspond to the grid size $dx = 0.05c/\omega_{pe}$, while the temporal resolution is chosen to be as $dt = 0.02\omega_{pe}^{-1}$.  The number of particles per cell is taken to be 8. From $x=0$ to $1000c/\omega_{pe}$, there is a vacuum region, and the plasma boundary starts from  $1000c/\omega_{pe}$. These have also been shown in Table \ref{Table}, along with the possible realistic values that this choice of parameters can correspond to.
The schematic of the simulation geometry is shown in Fig. \ref{Fig:schematic}. The laser pulse propagation is in the $\hat{x}$-direction. The electric and magnetic field components of the laser pulse are in $\hat{y}$ and $\hat{z}$-direction, respectively. The applied magnetic field is also along the $\hat{z}$ direction. It is chosen to be high at the plasma edge, as depicted in the schematic, and falls linearly with $x$. This geometry essentially supports the extraordinary EM modes. Even though the plasma is overdense, the presence of the externally applied magnetic field permits the propagation of electromagnetic wave if its frequency lies in the pass band of the dispersion curve depicted in Fig. \ref{Fig:dispersion_curve}. The (a) and (b) subplots have been drawn for the magnetic field values $B_0 = 10$ and $B_0 = 2.4$. The chosen fundamental laser frequency and its second harmonic lie in the pass band of the dispersion curve for the case of $B_0 = 10$. For $B_0 = 2.4$, the fundamental laser frequency hits the LH resonance. The second harmonic lies in the stop band. Thus, the external magnetic field allows the laser to penetrate inside the bulk of the target.  Subsequently,  as the magnetic field decreases and hits the value of $2.4$, the fundamental frequency hits the stop band. The second harmonic is observed to be generated at this location. However, it is also unable to propagate deeper inside the target as it lies in the stop band.

\begin{figure*}
  \centering
  \includegraphics[scale = 0.26]{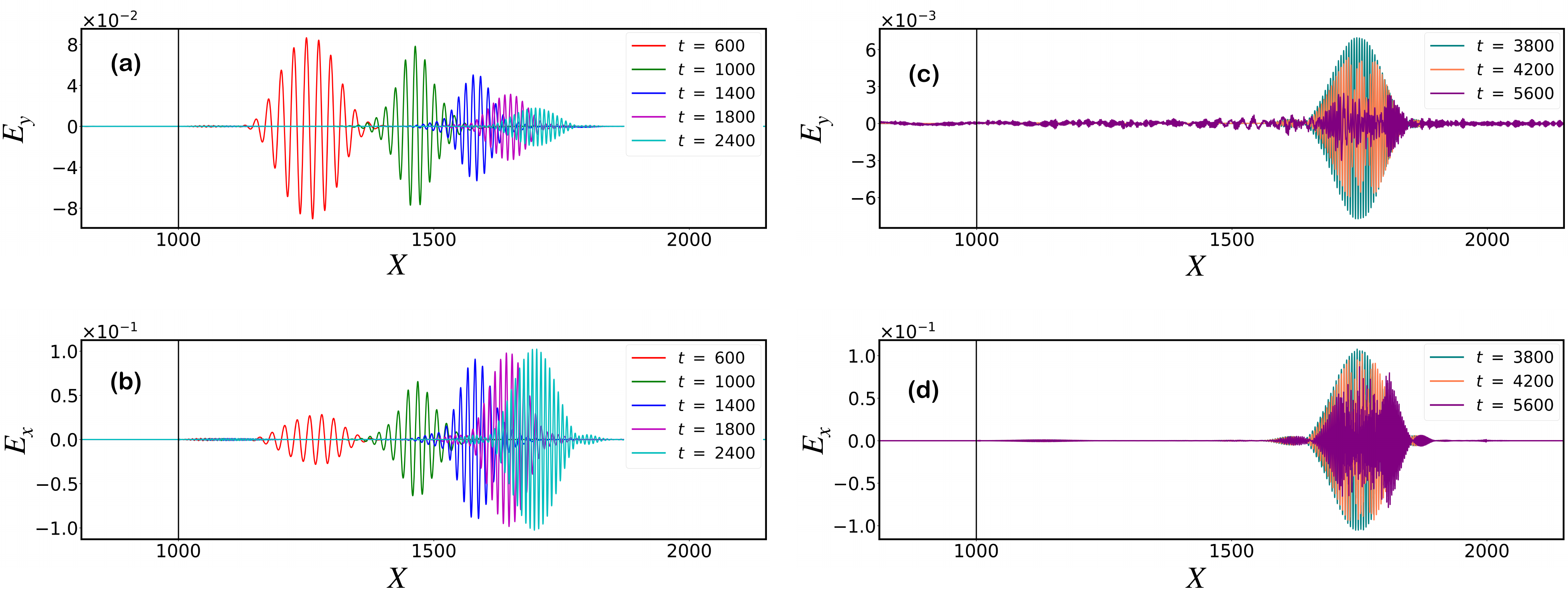}
  \caption{The $y$ and $x$ components of the electric field ($E_y$ and $E_x$) versus $x$ have been shown at different times of the simulation runs in (a),(b),(c), and (d), respectively. Here, the different colors have shown the $x$ locations of the fields at different times of the simulation run ($t$ = $600$ to $5600$).}
\label{Fig:EyEx_evolution}
\end{figure*}

\begin{figure*}
  \centering
  \includegraphics[scale = 0.25]{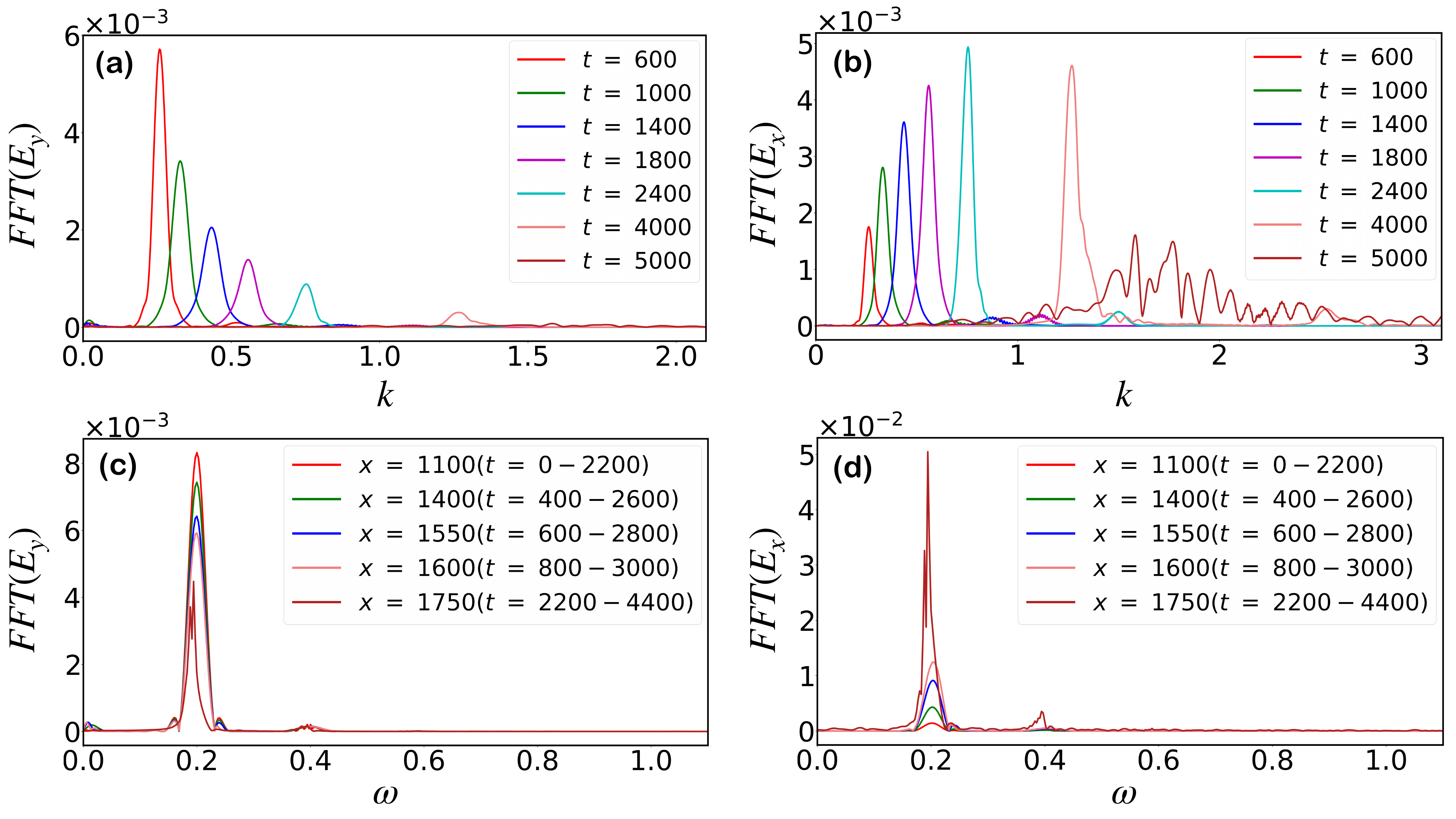}
  \caption{FFT of $E_y$ and $E_x$ with $k$ at different times have been shown by various colored lines in (a) and (b), respectively. FFT of $E_y$ and $E_x$ with $\omega$ at different $x$ locations have been shown in (c) and (d), respectively.}
\label{Fig:FFTs}
\end{figure*}

\begin{figure}
  \centering
  \includegraphics[scale = 0.22]{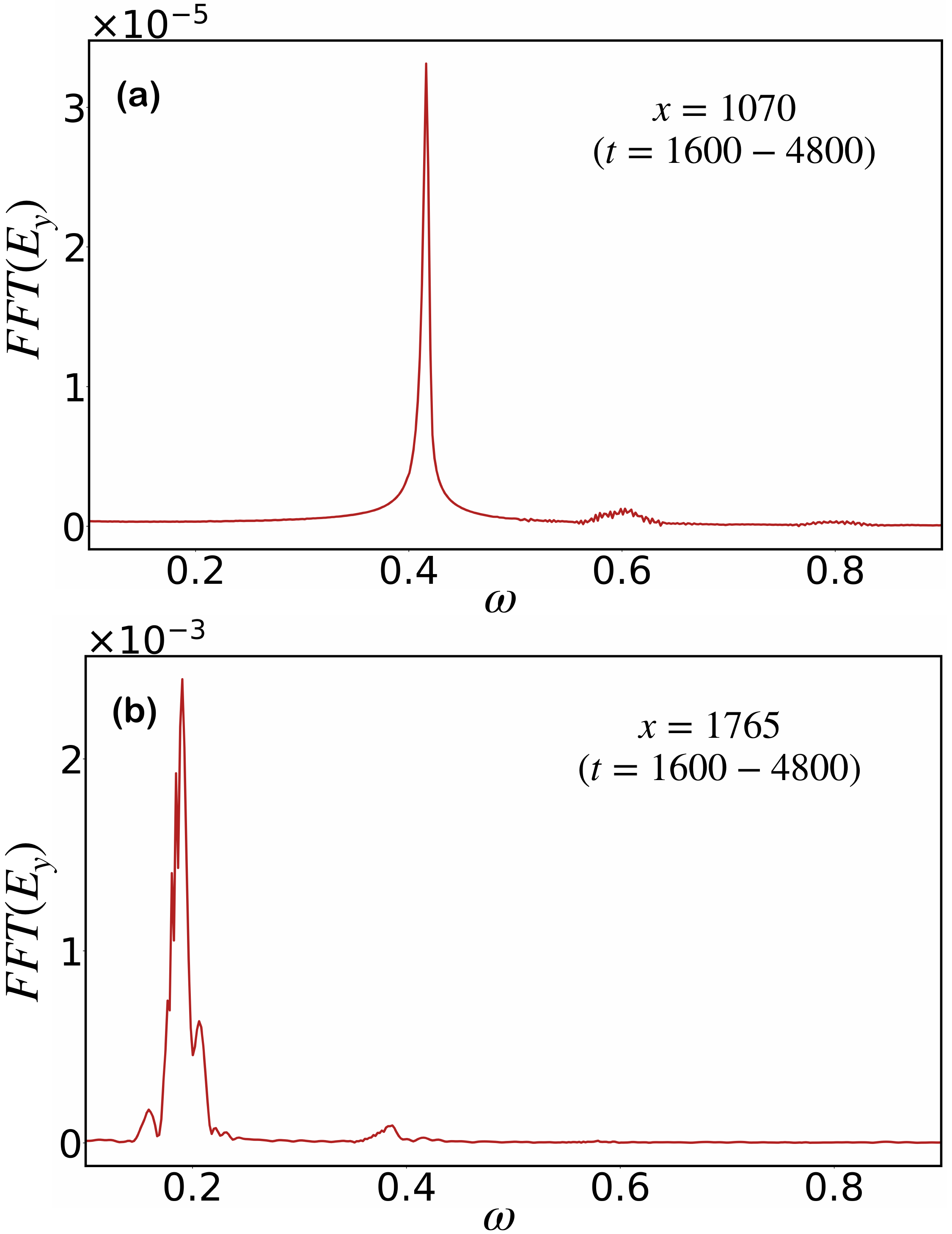}
  \caption{FFT of $E_y$ with $\omega$ showing generation of harmonics at two locations (a) $x=1070$, and (b) $x=1765$.}
\label{Fig:HHG}
\end{figure}

\begin{figure}
  \centering
  \includegraphics[scale = 0.13]{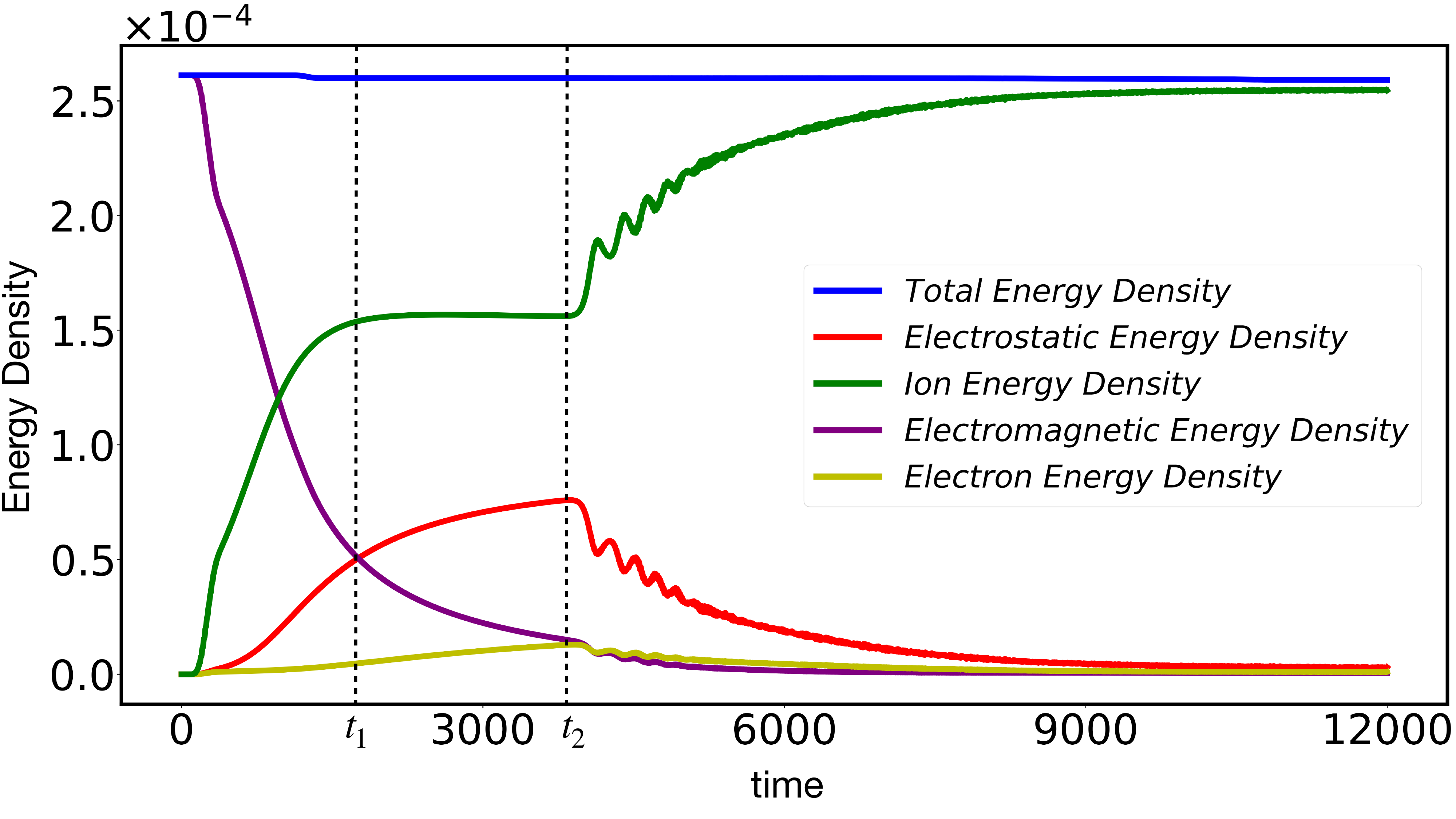}
  \caption{Time evolution of spatially averaged Ion, Electron, Electromagnetic, Electrostatic, and total energy density.}
\label{Fig:EnergyEvolution}
\end{figure}

\begin{figure}
  \centering
  \includegraphics[scale = 0.13]{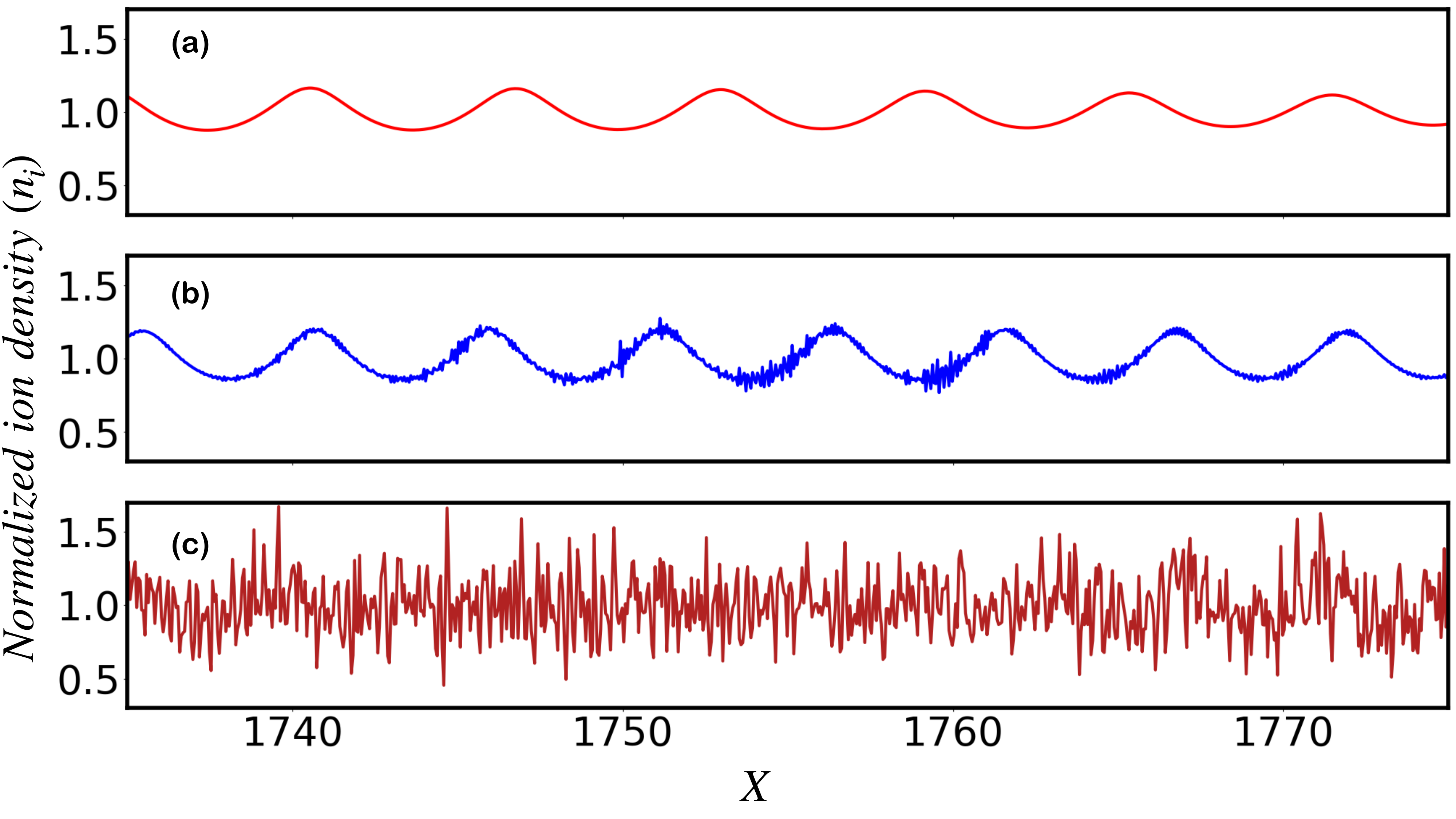}
  \caption{Variation of Ion density around resonance point at (a) $t=3200$, (b) $t=3800$, and (c) $t=4400$, before and after the wave-breaking.}
\label{Fig:Density}
\end{figure}

\begin{figure}
  \centering
  \includegraphics[scale = 0.23]{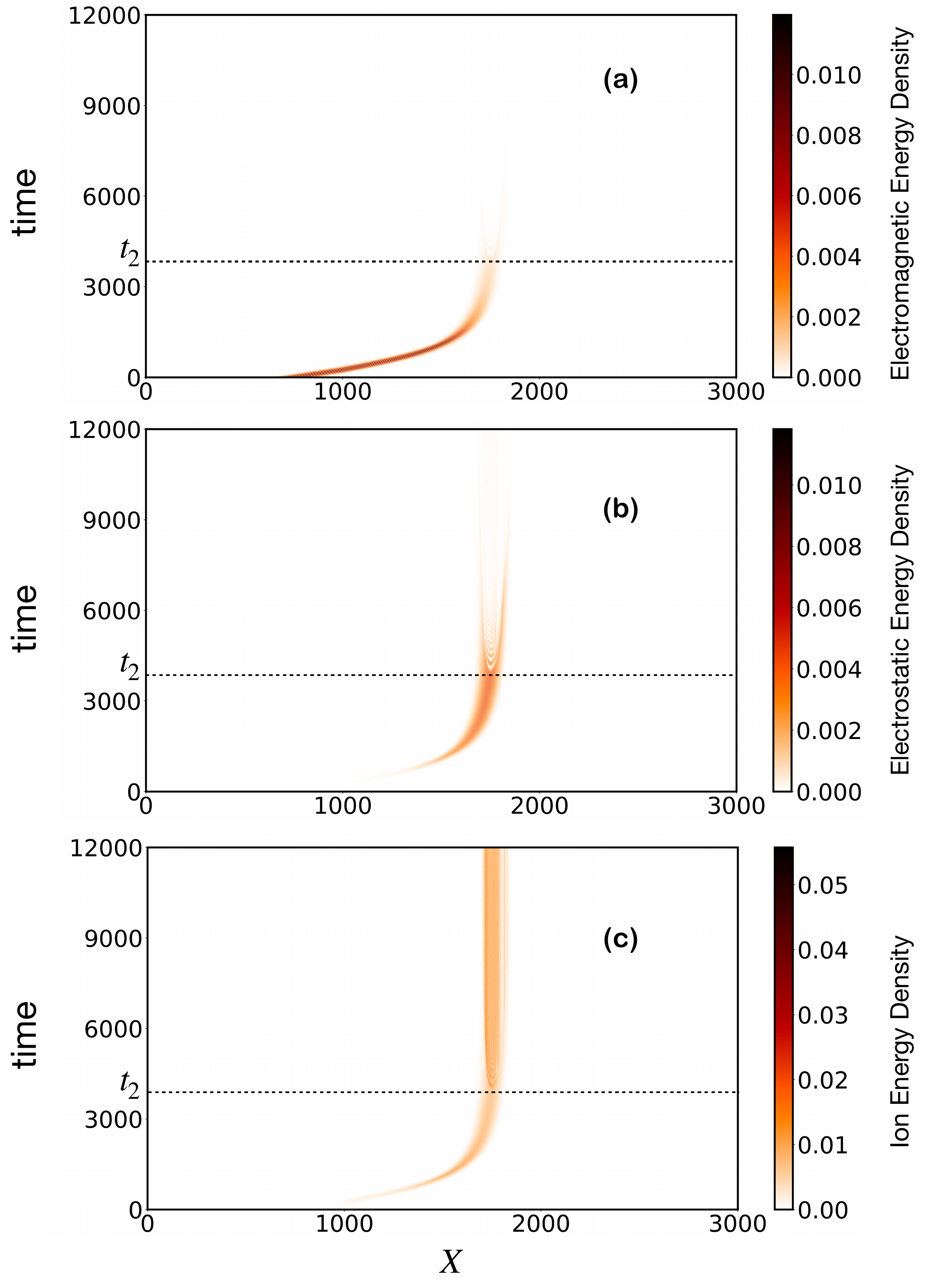}
  \caption{Variation of (a) Electromagnetic energy density, (b) Electrostatic energy density, and (c) Ion energy density, showing localized heating in the bulk of the plasma.}
\label{Fig:SpaceTimeEnergy}
\end{figure}

  \begin{figure}
  \centering
  \includegraphics[scale = 0.13]{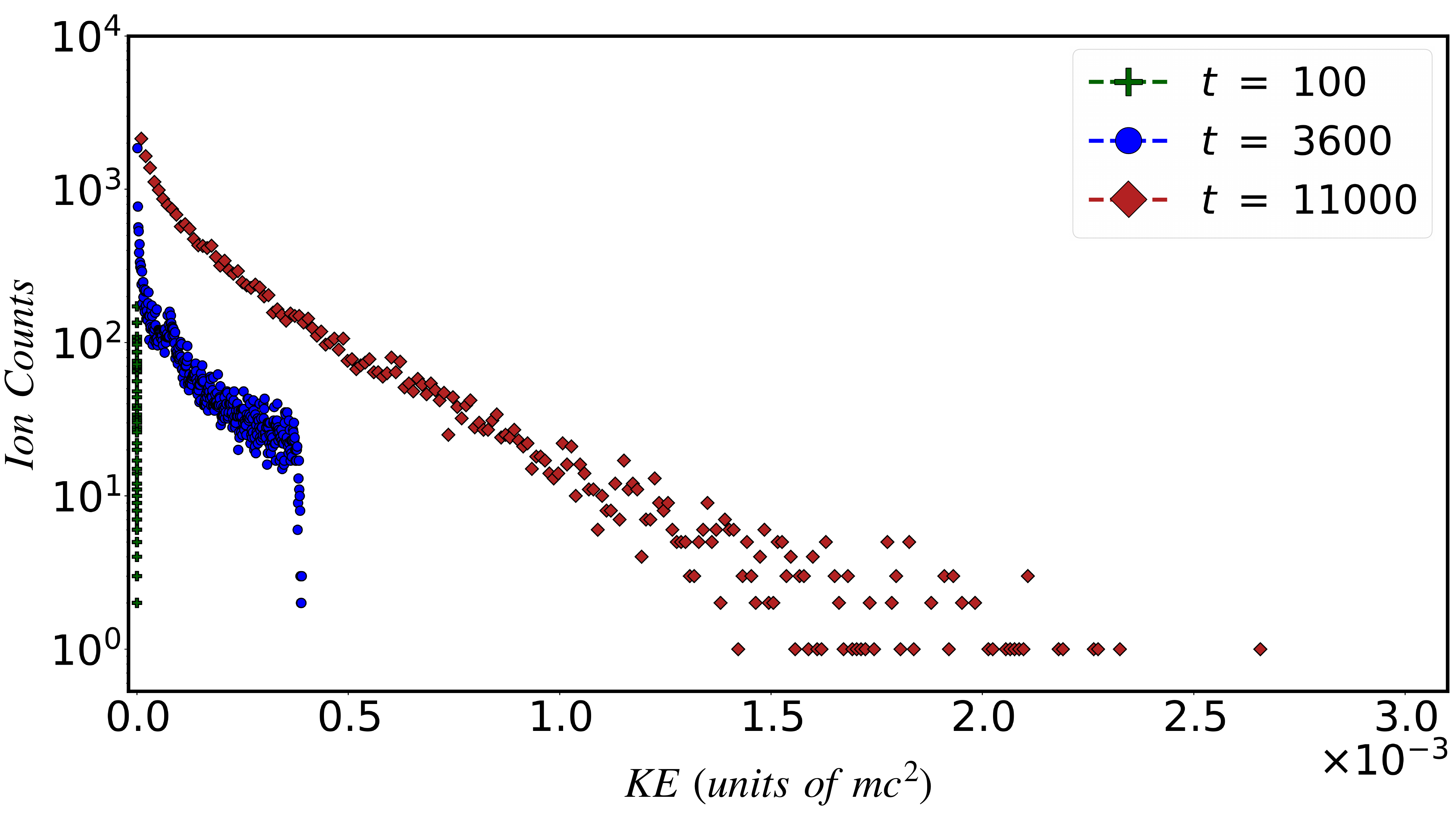}
  \caption{Ion counts vs Kinetic energy of ions at various time steps showing higher energy absorption after wave-breaking.}
\label{Fig:KE}
\end{figure}

\section{Observations} \label{sec:ResultDiscussions}

In this section, the simulation results are presented along with  its understanding.   
The incident laser pulse is plane polarised with its electric field along the $\hat{y}$ direction. The plasma is overdense for the chosen laser frequency. However, the laser frequency lies in the pass band of the X mode and hence propagates inside the plasma. This is shown in Fig. \ref{Fig:EyEx_evolution}(a) and Fig. \ref{Fig:EyEx_evolution}(c), which show the snapshots of $E_y$ at various times. As the pulse encounters an inhomogeneous magnetic field inside the plasma, the amplitude of the $E_y$ component is observed to diminish. However, concomitantly, there is a growth of $E_x$ (the $\hat{x}$ component of the electric field ) as seen from the subplot (b) of the same figure. Thus, the electrostatic character of the wave keeps increasing. Ultimately, the laser pulse encounters the LH resonance layer, and the EM wave propagation stops completely, with group velocity being zero at the resonance layer. Both the electrostatic and the electromagnetic pulses (e.g. $E_x$ and $E_y$ is this geometry) are subsequently observed to suffer wave-breaking (see $t= 4200 $ in Fig. \ref{Fig:EyEx_evolution}(c), and Fig. \ref{Fig:EyEx_evolution}(d)). The generation of shorter scales is evident from the figure. We also have Fourier analyzed these fields to show the formation of short-scale spatial structures with time in Fig. \ref{Fig:FFTs}. The time FFT at this location also shows the generation of the second harmonic. However, since the second harmonic is in the stop band for $B_ 0 = 2.4$, it also does not propagate further inside. 

The emission of higher harmonic and their propagation details can be observed from Fig. \ref{Fig:HHG}. We have considered two spatial locations, $x=1070$, before the resonance layer and at $x=1765$, around the resonance layer where the mode has stopped. The fundamental frequency of the mode is the same as that of the incident laser pulse, i.e., $0.2$. At the location, $x=1070$, second, third, and fourth harmonics are observed with efficiencies in descending order, but no fundamental frequency is observed. This location is far from the resonance layer on the left side. 
This shows that there is a generation of harmonics of the fundamental frequency from the resonance point by the EM pulse. The fundamental mode does not get reflected but the harmonics that get generated get transmitted in the backward direction to reach this point. In contrast, at the location $x=1765$, which is around the resonance layer, the presence of fundamental as well as second harmonic is clearly observed.  No signal was observed from the locations which are at the right-hand side of the resonance where the magnetic field strength becomes considerably weak to permit any propagation of the signals, be it fundamental and/or the harmonics that get generated at resonance. 

In  Fig. \ref{Fig:EnergyEvolution}, we show the evolution of various energies over time. The $x$-axis represents time, while the $y$-axis represents the energy density. The total energy shown by the blue-colored solid line is conserved. There is, however, a small dip at $ t = 1350$, which occurs when the reflected pulse from the plasma vacuum boundary moves out of the left boundary. 
Initially, all the energy is in electromagnetic fields and is shown by 
violet-colored solid line.    However, as time progresses, the electromagnetic energy gradually decreases. A simultaneous increase in electrostatic energy accompanies this decrease. The kinetic energy of the particles (shown by the green-colored solid line) also increases during this period. The increase in ion energy is rapid compared to the increase in the energy of electrons.  At time $t_1 = 1700$, the laser pulse hits the resonance layer and comes to a standstill. From $t_1$ to $t_2$, there is a steady conversion of electromagnetic field energy to electrostatic fields. The wave profile in this duration remains intact (see Fig. \ref{Fig:EyEx_evolution}). No ion energy increase is observed during this period.  At $t_2 = 3800$, the electrostatic field acquires sufficient magnitude to undergo wave-breaking phenomena (See Fig. \ref{Fig:EyEx_evolution}(d)). 
Thus, for  $t > t_2$, the pulse suffers wave-breaking, which leads to decay in both electrostatic and electromagnetic energies, as can be observed in Fig. \ref{Fig:EnergyEvolution}. The green solid curve shows that energy is now being transferred to ions. 

The evidence of wave-breaking is provided in Fig. \ref{Fig:Density}, where the ion density around the LH resonance point as a function of space has been shown for three distinct times.  At $t=3200$, when the waveform is intact at the resonance layer, the ion density has a nice sinusoidal form. At $t=3800$, when the wave starts to break, ion density in the vicinity of resonance shows the onset of short-scale generation, and the sinusoidal form begins to get disturbed. At a later time, $t=4400$, the waveform gets broken, and random fluctuations appear, which show spiky characteristics in ion density.

There is a strong transfer of field energy to the particles after the wave breaks. So far the electrostatic energy had been steadily rising.  It now starts to decay. Electromagnetic energy continues to decay. The ion kinetic energy that had plateaued after $t_1$ now starts to increase. The electrostatic energy here provides a conduit for transferring field energy to particles at this particular time. 
This is also evident from Fig. \ref{Fig:SpaceTimeEnergy}, where the spatially integrated electromagnetic, electrostatic, and ion kinetic energy has been plotted as a function of time. It can be seen that the electromagnetic energy steadily decreases. The electrostatic energy and the kinetic energy increase together for some duration. Thereafter, after  $t_2 = 3800$, when wave-breaking starts, the electrostatic energy diminishes, and the kinetic energy of the ion picks up simultaneously.   

\begin{figure}
  \centering
  \includegraphics[scale = 0.13]{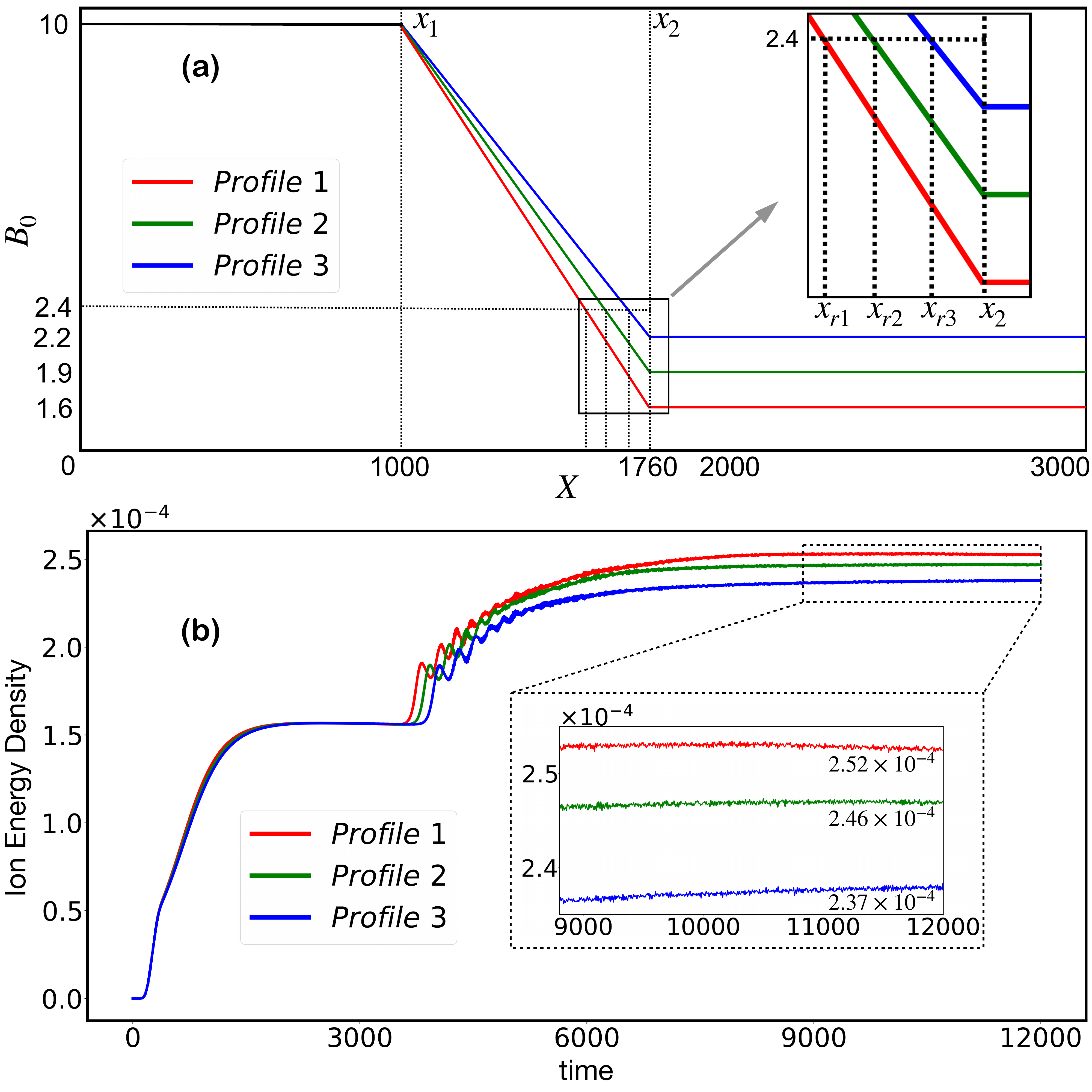}
  \caption{Various magnetic field profiles shown in (a) and the time evolution of Ion energy density with these profiles shown in (b).}
\label{Fig:profile}
\end{figure}

\begin{figure*}
  \centering
  \includegraphics[scale = 0.25]{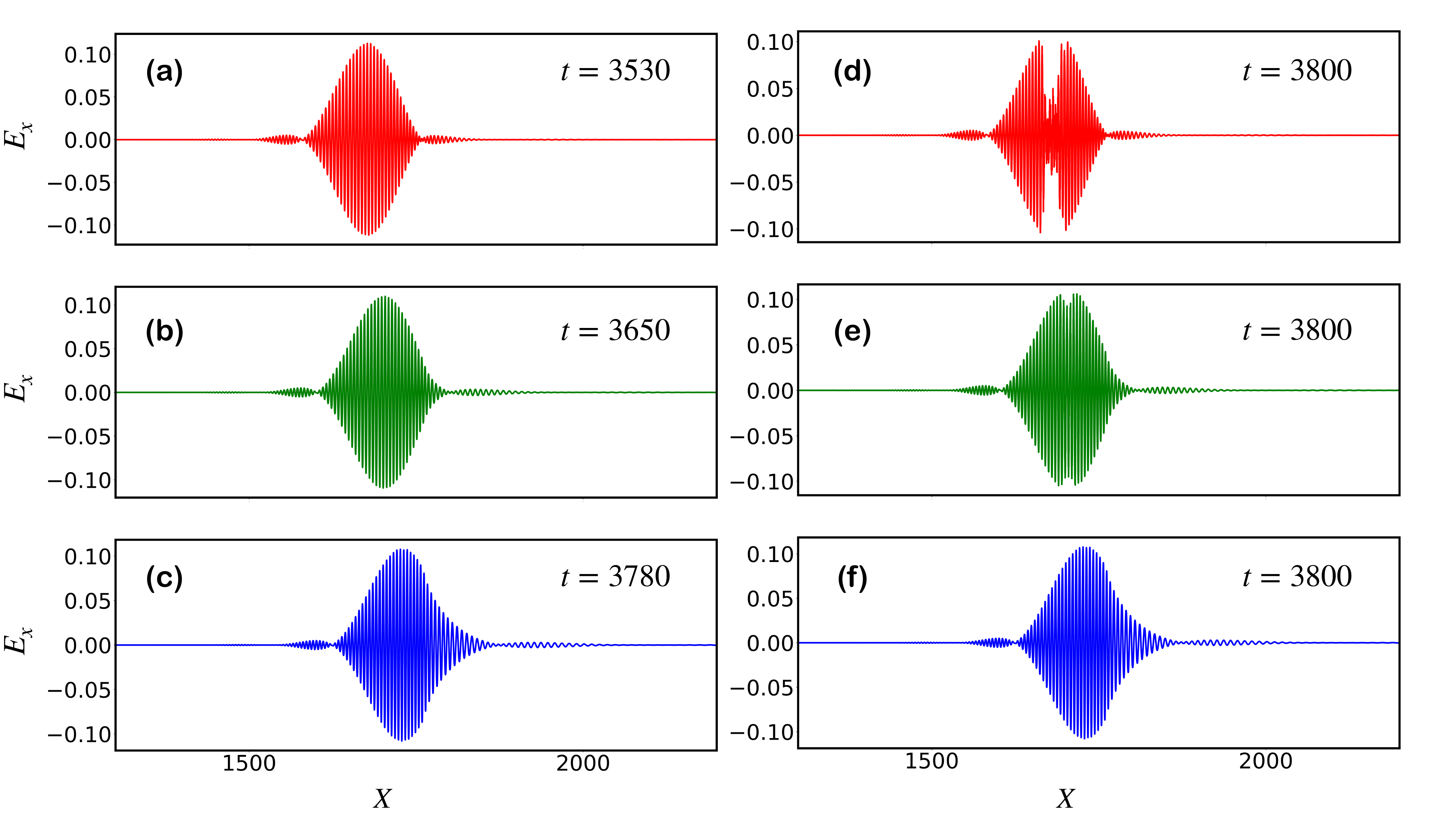}
  \caption{$E_x$ vs $x$ at various times for (a) Profile 1, (b) Profile 2, and (c) Profile 3. $E_x$ vs $x$ at time $t = 3800$ for (d) Profile 1, (e) Profile 2, and (f) Profile 3. }
\label{Fig:ExwithProfile}
\end{figure*}

\begin{figure}
  \centering
  \includegraphics[scale = 0.13]{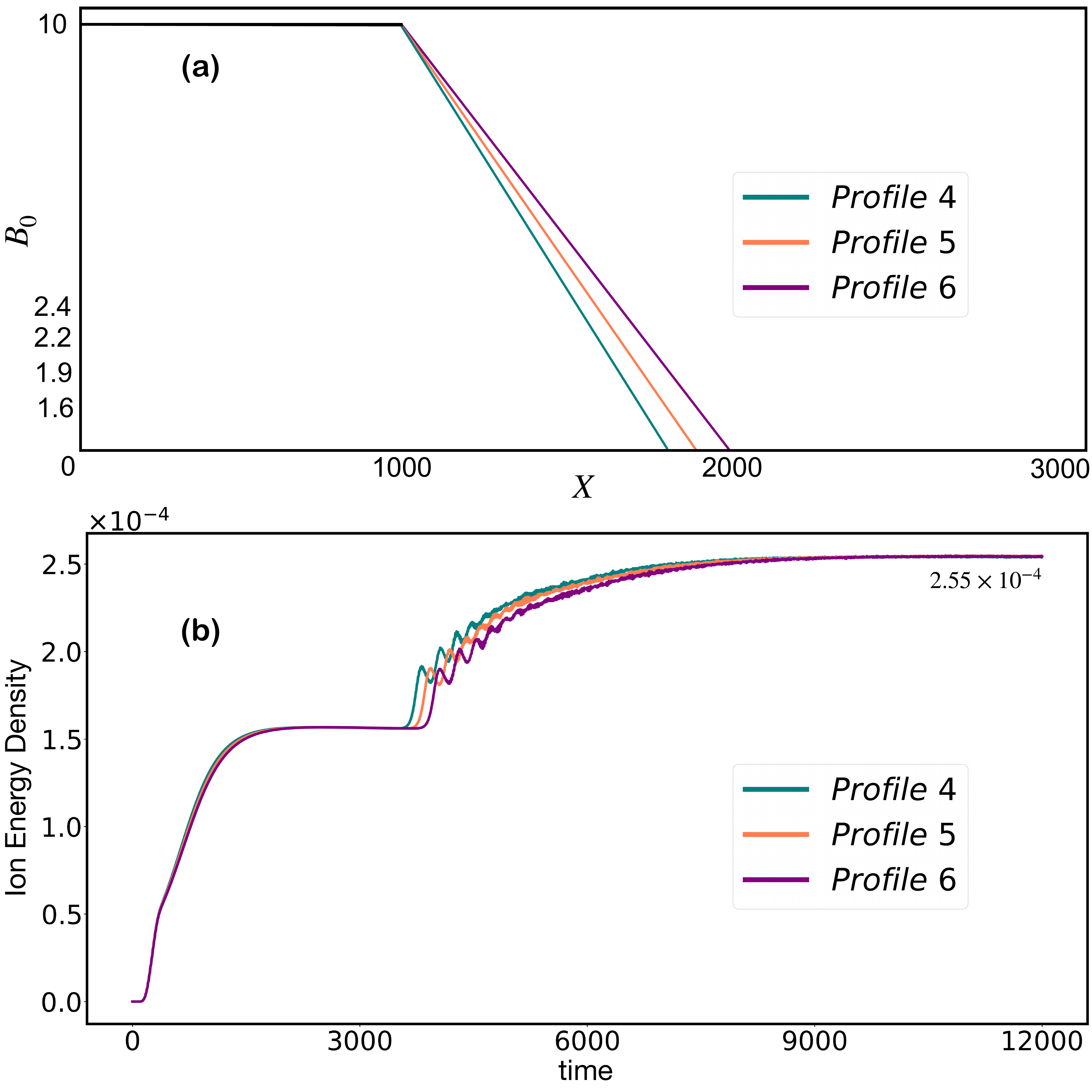}
  \caption{Various magnetic field profiles shown in (a) and the time evolution of Ion energy density with these profiles shown in (b).}
\label{Fig:profileExtended}
\end{figure}
 
In Fig. \ref{Fig:KE}, we have shown the ion counts as a function of its kinetic energy. There is a significant increase in ion energy after the wave-breaking process. 
Our studies thus demonstrate that by suitably choosing the spatial profile of the magnetic field, the EM wave can be enabled to enter the bulk region of the plasma, and then it can be made to dump its energy to ions at a suitable location. 

We now study the possible role, other than the resonance location,  the choice of the spatial profile of the magnetic field plays in the energy absorption process. 
In Fig. \ref{Fig:profile}(a),  three choices of various magnetic field profiles have been shown.  All three profiles start with $B_0 = 10$. Thereafter, from  $x = 1000$, they decay linearly with different slopes till $x = 1760$; thereafter, the value of the magnetic field is again a constant. However, this constant magnetic field differs for the three profiles, as shown in Fig. \ref{Fig:profile}(a). 
 The energy acquired by the ions for each of these profiles has been shown alongside in Fig. \ref{Fig:profile}(b). It is observed that for the steeper magnetic field profile, the $x$ location happens to be ahead compared to the resonance points defined for other cases. Thus, wave-breaking happens at an earlier time. It should also be noted 
 that this steeper magnetic field profile helps in the higher absorption of laser energy by ions. 

\begin{figure*}
  \centering
  \includegraphics[scale = 0.25]{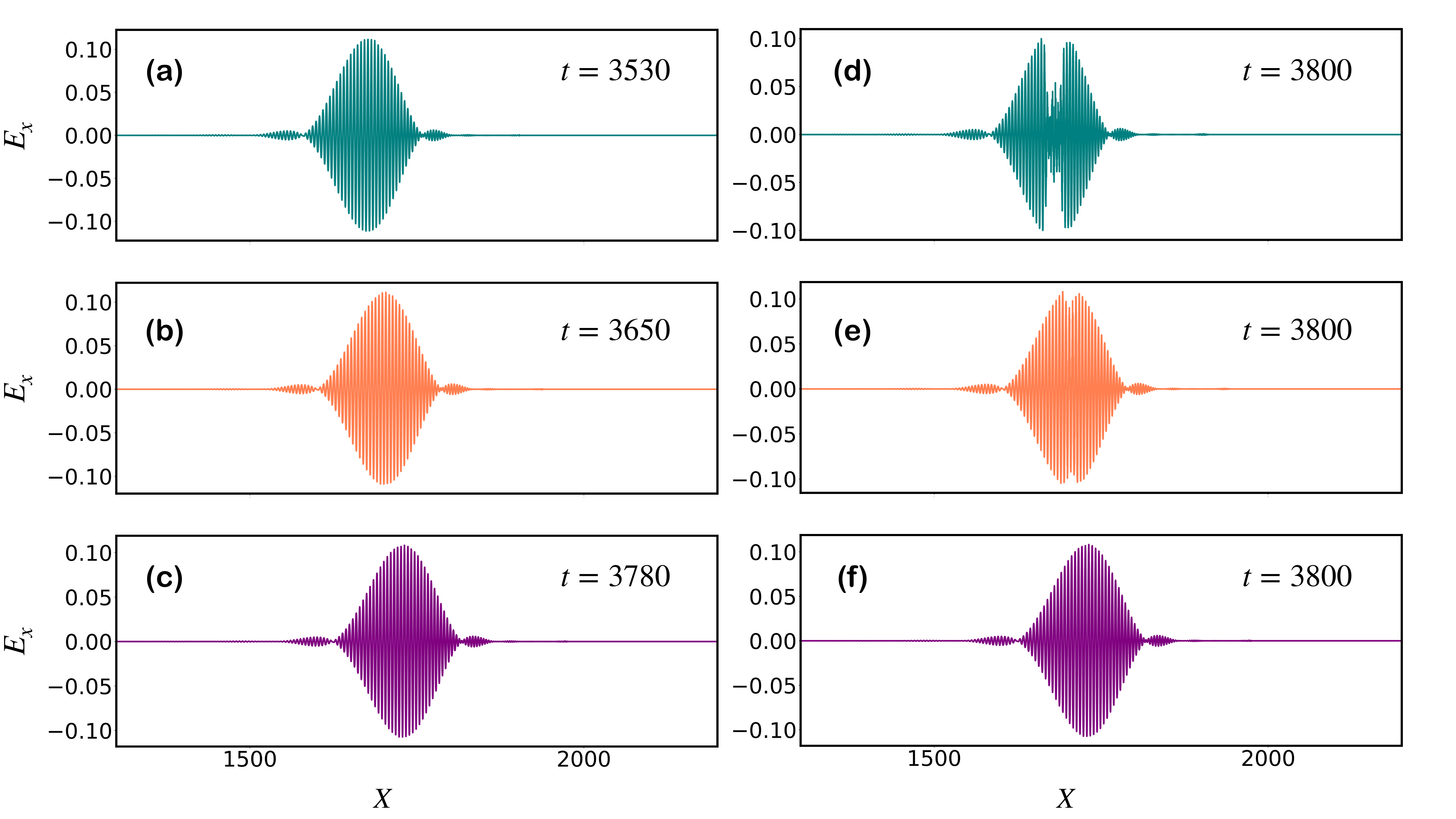}
  \caption{$E_x$ vs $x$ at various times for (a) Profile $4$, (b) Profile $5$, and (c) Profile $6$. $E_x$ vs $x$ at time $t = 3800$ for (d) Profile $4$, (e) Profile $5$, and (f) Profile $6$.}
\label{Fig:ExwithProfileExtended}
\end{figure*}

A comparison of the wave pulse profile at times just before the onset of wave-breaking phenomena in each of the cases has been depicted in Fig. \ref{Fig:ExwithProfile}. The plots in the right column are compared at the same time. For the choice of $B_0$ profile ($1$) and ($2$), the EM wave pulse has already started degenerating. It is clearly faster for profile ($1$). 
For the choice of shallower $B_0$ field profile ($3$), the wave-breaking has not yet started. A comparison of subplots (c) and (f) shows that there is hardly any difference between them, though the pulse has been depicted at different times. Both the times are, however, before the onset of wave-breaking. Thus, the wave preserves its shape, standing at the resonance point for a significant time. It should also be noted that the shallower magnetic field profile shown in subplot (c) seems to be stretched toward the right-hand side compared to the other two cases shown in subplot (a) and (b). 
To ascertain whether this happens as a result of the (i) difference in slope or the (ii) fact that for profile ($3$), the spatial resonance point is near compared to the spatial location $x = 1760$ where the magnetic field again attains a uniform value, we have carried out another set of simulations with the $B_0$ field profiles depicted in Fig. \ref{Fig:profileExtended}(a). In this set of simulations, the three slopes of the $B_0$ field are identical to the previous choice; however, for each of these cases (profiles $4$, $5$, and $6$), ultimately, the magnetic field value goes to zero. The pulse profile for this set of choices has been shown in Fig. \ref{Fig:ExwithProfileExtended}, and it is clear that, in this case, none of the EM pulses show a stretched form. It should thus be realized that in the former set of simulations, the stretching of the EM wave profile for the profile ($3$) happens as 
there is a very small difference between the resonance value of $B_0 = 2.4 $ and the lower constant value of $B_0 = 2.2$ and the $\Delta x = x_2 - x_{r3} = 19.48$ quite small. 
The stretched pulse indicates that it is, in a way, trying to tunnel through. 

\begin{figure}
  \centering
  \includegraphics[scale = 0.13]{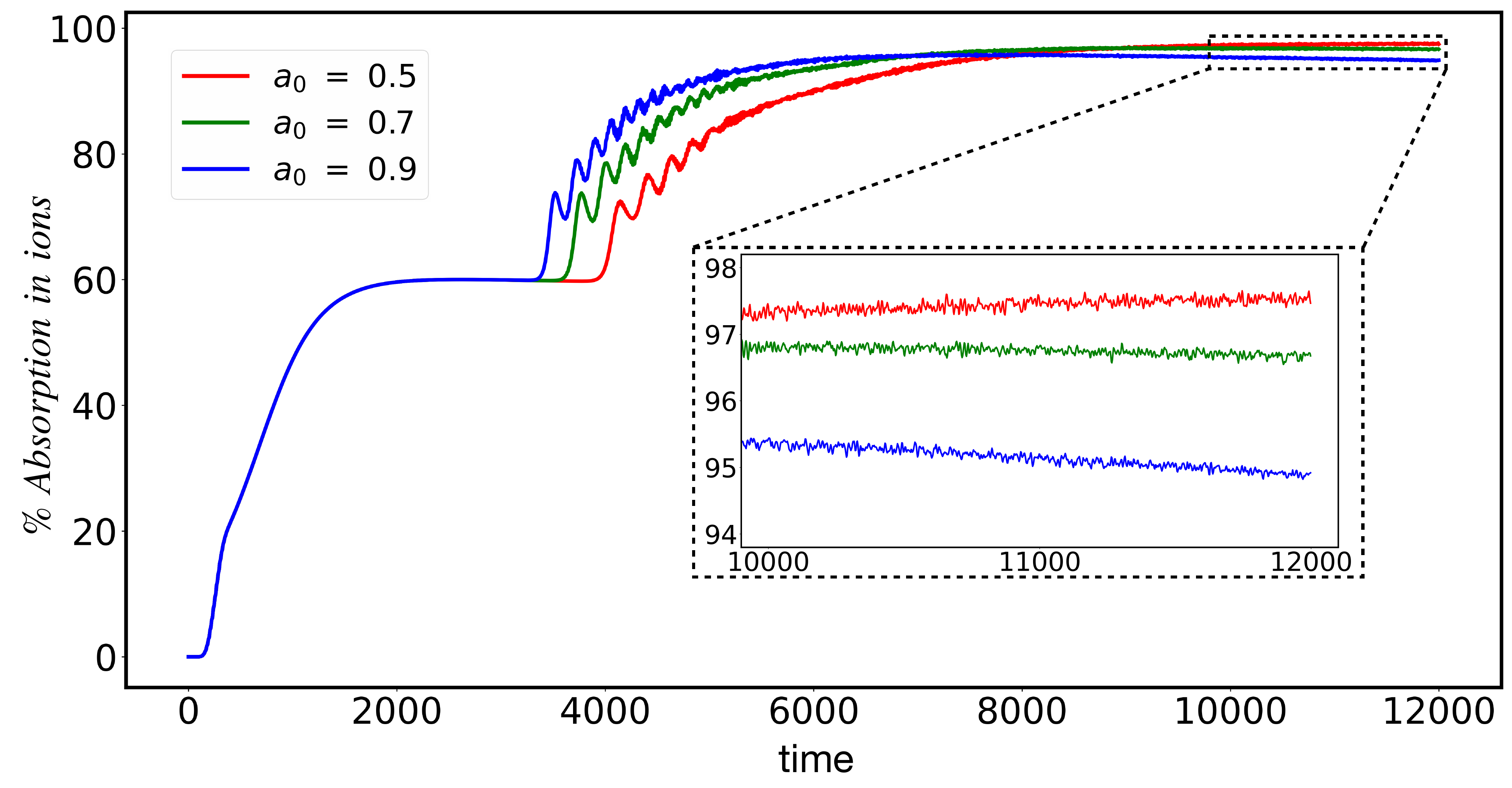}
  \caption{Time evolution of percentage energy absorption by Ion, showing variation in the absorption with different $a_0$ values in the non-relativistic case.}
\label{Fig:intensity}
\end{figure}

\begin{figure}
  \centering
  \includegraphics[scale = 0.13]{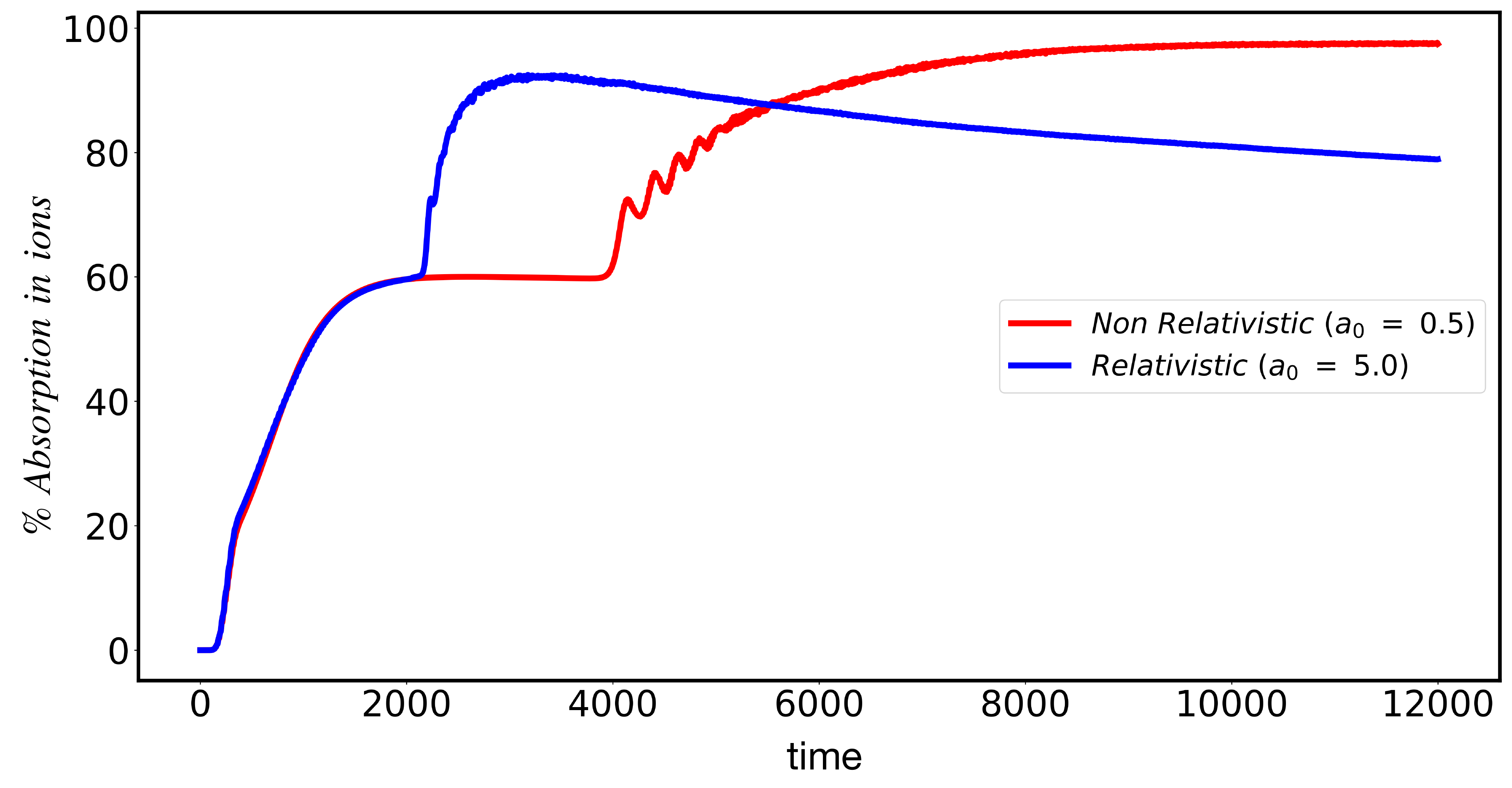}
  \caption{Comparison of Ion energy absorption between non-relativistic and relativistic case.}
\label{Fig:relativistic}
\end{figure}

\begin{figure}
  \centering
  \includegraphics[scale = 0.22]{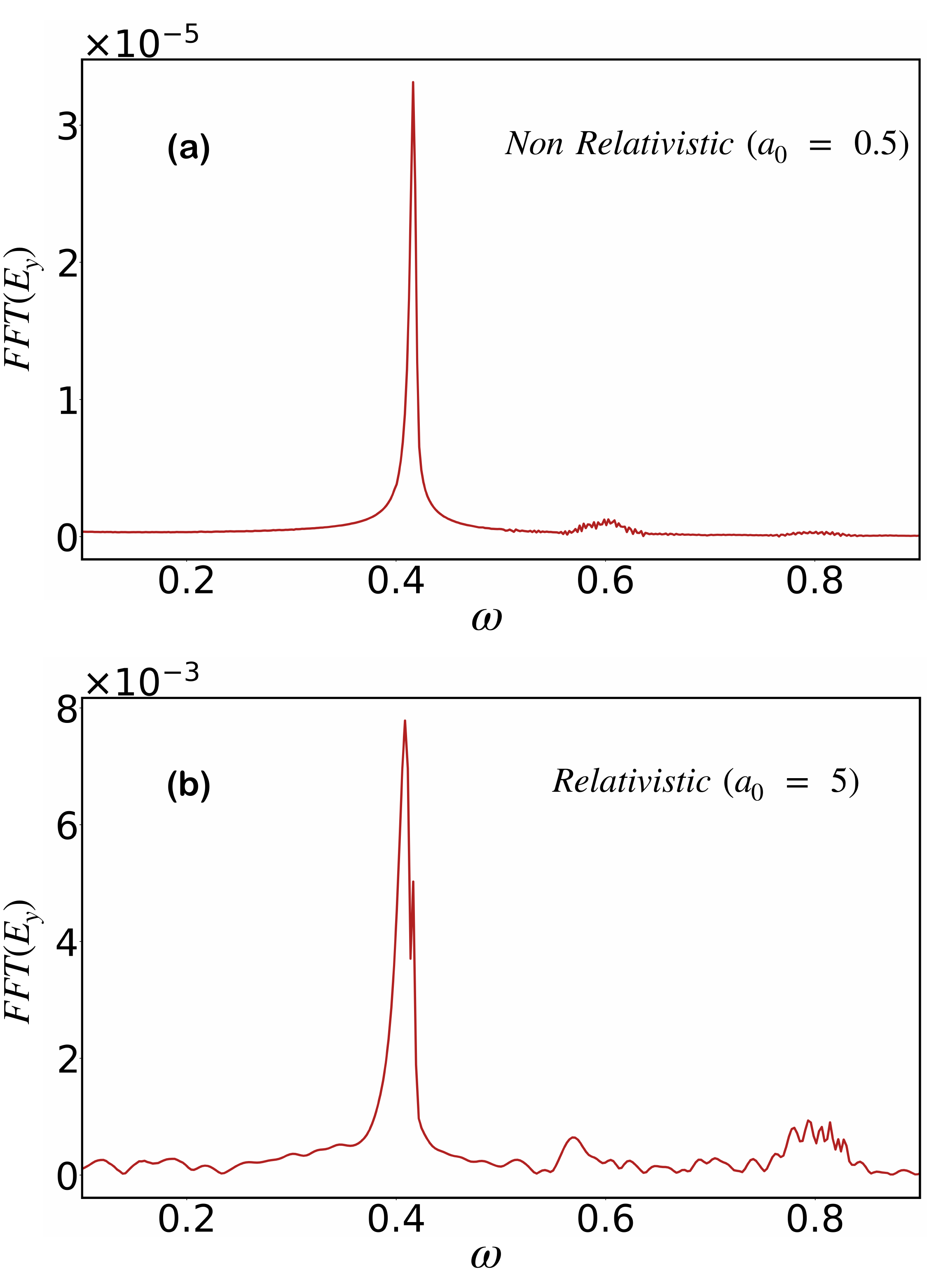}
  \caption{FFT of $E_y$ with $\omega$ showing generation of harmonics in (a) Non-relativistic, and in (b) Relativistic case}
\label{Fig:HHG_relativistic}
\end{figure}

\begin{figure}
  \centering
  \includegraphics[scale = 0.13]{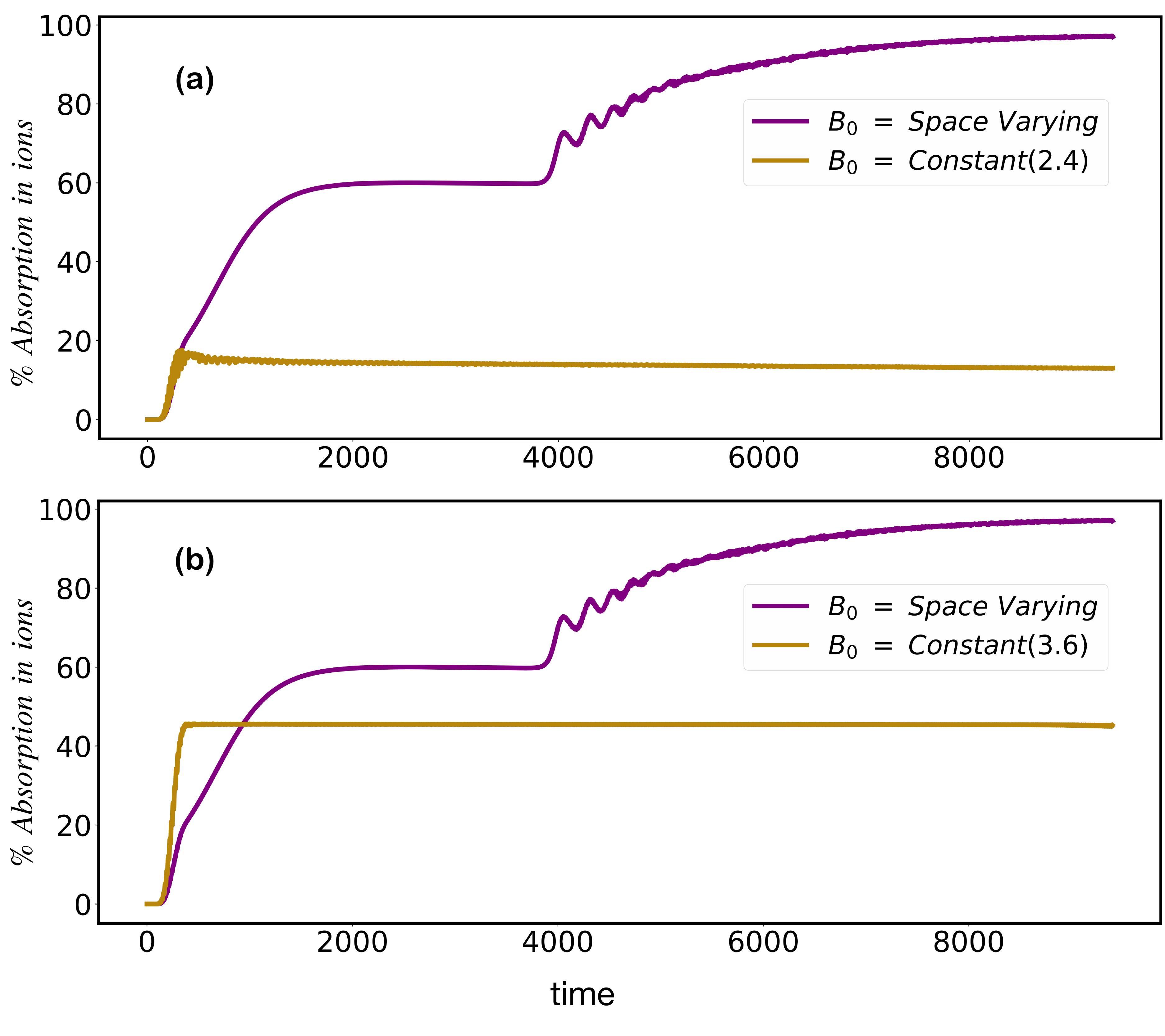}
  \caption{Percentage absorption in Ion for Space varying vs Constant external magnetic field (a) Constant $B_0 = 2.4$, and (b) Constant $B_0 = 3.6$}
\label{Fig:EnhancedAbs}
\end{figure}

Another feature to note is that for this particular set of simulations, the ultimate energy acquired by all three profiles is the same (See Fig. \ref{Fig:ExwithProfileExtended}(b)) and comparatively higher than the previous set of simulations. Thus, the slope of the magnetic field has a limited role. The resonance location is determined by the same. 
The absorption, on the other hand, is governed by the final value of the magnetic field $B_0$. 

We have also studied the influence of laser intensity on the absorption process. In Fig. \ref{Fig:intensity}, we have chosen three distinct choices of the parameter $a_0$. It is interesting to note that the onset of wave-breaking phenomena occurs earlier for the case with a higher value of $a_0$. This is understandable as the electrostatic wave will reach the wave-breaking threshold earlier with higher energy available from the EM wave pulse.  
The percentage absorption is, however, higher for lower-intensity cases. This comparison is even more dramatic when $a_0 = 5$ (a highly relativistic intensity) is chosen (see Fig. \ref{Fig:relativistic}). The explanation for this lies in the higher efficiency of the generation of harmonics when the laser intensity is higher. The harmonics reflect back from the resonance layer and escape the plasma. The FFT spectrum of the pulse near resonance in Fig. \ref{Fig:HHG_relativistic} shows the strength of the harmonics. It is clear that the intensity of harmonic is much higher for $a_0 = 5$ compared to $a_0 = 0.5$. 

Finally, in Fig. \ref{Fig:EnhancedAbs}, we provide a comparison of the percentage energy absorption that was achieved for the homogeneous case described in earlier publication (\cite{Juneja_2023}). In subplot (a), we show the case for which the homogenous case corresponds to $B_0 = 2.4 $, the resonance value of the magnetic field. In subplot (b), the optimized strength of the homogeneous $B_0$ field, which is near resonance, has been chosen. The optimization of energy absorption in the context of a homogeneous $B_0$ field was described earlier in (\cite{Juneja_2023}). The value should be close to resonance but not quite the resonance value to ensure that the EM wave frequency lies in the pass band and, at the same time, it is not too far away from the resonant value. By choosing an inhomogeneous magnetic field, one has been able to get rid of the shackles of satisfying two considerations, namely ensuring that resonance occurs and, at the same time, the EM pulse penetrates sufficiently inside the plasma, simultaneously with the same strength of the magnetic field. It is clear here that the percentage absorption in the inhomogeneous case is much higher than in the homogeneous case. 

\section{Conclusion\label{conclusion}}
Particle-In-Cell (PIC)  studies on laser/EM wave interaction with a magnetized plasma have been carried out to investigate the phenomena of laser energy absorption by plasma ions.  Previous studies (\cite{vashistha2020new,Juneja_2023}) in this direction have shown the possibility of ion heating in the lower Hybrid regime of operation. However, a major restriction was choosing a possible magnetic field that would ensure both the occurrence of Lower hybrid resonance and propagation of the laser/EM pulse inside the plasma. Clearly, both conditions cannot be simultaneously satisfied as the group speed of the EM pulse vanishes at the resonance. One, therefore had to choose an optimized value of the $B_0$  field, which is not exactly at the resonance but ensures the EM wave pulse to be in the pass band of the dispersion curve to yield the best absorption of energy. By choosing an inhomogeneous magnetic field this lacunae has been overcome. The $B_0$ field strength is chosen to be higher at the plasma edge to ensure smooth propagation of the laser/EM ave pulse inside the plasma. The strength of the magnetic field is decreased in space subsequently so that the pulse reaches the resonance point within the plasma. 
The simulation clearly shows that the laser energy gets dumped at the localized spot where the resonance condition is met. Furthermore, the efficiency of laser energy absorption improves significantly as the pulse meets the exact resonance condition. 
The role of the spatial magnetic field profile, laser intensity, etc., have also been studied in detail.

\section*{Acknowledgements}
The authors would like to acknowledge the OSIRIS Consortium, consisting of UCLA and IST (Lisbon, Portugal), for providing access to the OSIRIS-4.0 framework, which is the work supported by the NSF ACI-1339893. AD would like to acknowledge her J C Bose fellowship grant JCB/2017/000055 and CRG/2022/002782 grant of DST. The authors thank the IIT Delhi HPC facility for computational resources. Rohit Juneja thanks the Council for Scientific and Industrial Research(Grant no. 09/086(1448)/2020-EMR-I) for funding the research.

\bibliographystyle{elsarticle-harv}
\bibliography{Absorption.bib}






\end{document}